\documentclass[oldversion,letter]{aa}
\usepackage{graphicx}
\usepackage{txfonts}

\begin{document}
\title{Evidence of magnetic field wrapping around penumbral filaments}
\author{J.M.~Borrero\inst{1} \and B.W.~Lites\inst{1} \and S.K.~Solanki\inst{2}}

\offprints{J.M.~Borrero}

\institute{High Altitude Observatory, National Center for Atmospheric Research\fnmsep\thanks{
The National Center for Atmospheric Research in sponsored by the National Science Foundation},
  Center Green Dr. CG-1, Boulder CO, 80301, USA\\ \email{borrero@ucar.edu,lites@ucar.edu}
  \and
  Max Planck Institut f\"ur Sonnensystemforshung, Max Planck Strasse 2, 37191 Katlenburg-Lindau,
  Germany\\ \email{solanki@mps.mpg.de}}

\date{}

\abstract{We employ high-spatial resolution spectropolarimetric observations from
the Solar Optical Telescope on-board the Hinode spacecraft to investigate the fine 
structure of the penumbral magnetic fields. The Stokes vector of two neutral iron lines
at 630 nm is inverted at every spatial pixel to retrieve the depth-dependence of
the magnetic field vector, line-of-sight velocity and thermodynamic parameters.
We show that the azimuthal angle of the magnetic field vector has opposite sign 
on both sides above the penumbral filaments. This is consistent with the wrapping 
of an inclined field around the horizontal filaments. The wrapping effect is 
stronger for filaments with larger horizontal extensions. In addition, we find that
the external magnetic field can penetrate into the intraspines, leading to
non-radial magnetic fields inside them. These findings shed 
some light on the controversial small-scale structure of the sunspot penumbra.
}

\titlerunning{Magnetic field wrapping in penumbral filaments}
\authorrunning{Borrero et al.}

\maketitle

\section{Introduction}

Our knowledge of the small-scale structure of magnetic features in the solar 
photosphere remains hindered by the limited spatial resolution we can achieve. 
A prominent example of this situation is the fine structure of the sunspot penumbra,
where a variety of different models compete to explain the observations (Solanki \&
Montavon 1993; Schlichenmaier et al. 1998; S\'anchez Almeida 2005; Spruit \& Scharmer 2006; Borrero 2007).
The degree of sophistication of these models is such that we can no longer
use moderate spatial resolution ($\sim 1$") observations to distinguish among them.

Many of these models assume the presence of two different magnetic components: one
in the form of a somewhat inclined ($\sim 40-50^{\circ}$ with respect
to the vertical) and strong ($\sim 2000$ G) magnetic field, and another in the form
of a weaker and more horizontal magnetic field. They are usually referred to as
spines and intraspines, respectively. Their presence has been repeatedly confirmed 
observationally (Lites et al. 1993; R\"uedi et al. 1998; Borrero et al. 2004,2005; 
Bellot Rubio et al. 2004; Bello Gonzalez et al. 2005; Langhans et al. 2005).

Some of these models identify the intraspines (horizontal and weak
magnetic field) with a horizontal flux tube ({\it uncombed model}), 
whereas in others it appears as the consequence of a field free gap
protruding from beneath ({\it gappy model}). Either way, spines are 
always assumed to avoid the intraspines by folding around them (Mart{\'\i}nez 
Pillet 2000). Although studies of the azimuthal variation 
of the Net Circular Polarization (Schlichenmaier et al. 2002; M\"uller et 
al. 2002; Borrero et al. 2007) and spectropolarimetric observations at
moderate spatial resolution (Lites et al. 1993) provide an indirect measure for 
this effect, direct observational evidence supporting this assumption has never been presented.

In this work we use Hinode's high-spectral ($\Delta \lambda / \lambda < 4\times10^{-6}$)
 and high-spatial ($\sim 0.32$") resolution polarimetric observations of the pair of 
photospheric Fe I lines at 630 nm to retrieve the depth dependence of the magnetic
 field vector. In section 2 we describe those observations. Section 3 presents the inversion procedure that
allows us to infer the magnetic field vector from the recorded polarized spectra.
Section 4 presents our results and finally, Section 5, the conclusions.

\section{Observations}

On May 3rd 2007, between 10:15 and 11:40 am UT, the active region AR 10953 was mapped using the spectropolarimeter of
the Solar Optical Telescope on-board of the Hinode spacecraft (Lites et al. 2001, Kosugi et al. 
2007; Shimizu et al. 2007). The active region was located at a heliocentric angle of $\theta=19.2^{\circ}$.
It was scanned in a thousand steps, with a step width of 0.148" and a
slit's width of 0.158". The spectropolarimeter recorded the full Stokes vector ($I$, $Q$, $U$ and $V$)
of the pair of neutral iron lines at 630 nm with a spectral sampling of 21.53 m\AA. 
The integration time was 4.8 seconds, resulting on an approximate noise level of $1.2 \times 10^{-3}$.
In the absence of the telluric oxygen lines we proceeded with two different wavelength calibration
methods that were cross-checked for consistency. The first method was obtained by matching the average
quiet Sun profile with the FTS spectrum, whereas the second calibration assumes that the average umbral
profile exhibits no velocities, which according to Rezaei et al. (2006) is a reasonable assumption.

Map of the continuum intensity on the limb-side at 630 nm is shown in Figure 1. The black
arrow indicates the direction of the center of the solar disk. The penumbra on the center 
side is heavily distorted and therefore left out from our analysis.
On the limb side the penumbra is more uniform, with radially aligned filaments. This 
is the region that we have chosen to study. This sunspot has negative polarity (magnetic field in the umbra points
towards the solar interior), however the results presented hereafter are shown, 
in order to facilitate the interpretation, as if the sunspot had positive polarity.

\begin{figure}
\begin{center}
\includegraphics[width=7cm]{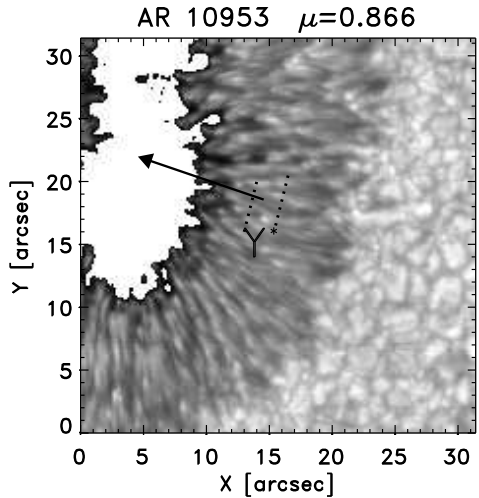}
\caption{Continuum intensity map at 630 nm of the limb-side of AR 10953. The umbra has been
removed to increase the image contrast. This sunspot
was observed using Hinode's spectropolarimeter on the 3rd of May, 2007 at an 
heliocentric angle of $\theta = 19^{\circ}$. The black arrow points towards the center
of the solar disk. The pair of dashed lines indicated by Y$^{*}$ lie near the line of symmetry
of the sunpot. Results along these two slices are discussed in Section 4.}
\end{center}
\end{figure}

\section{Inversion procedure}

To retrieve the physical parameters of the solar atmosphere from the spectropolarimetric
observations, we employ the SIR code (Ruiz Cobo \& del Toro Iniesta 1992).
This code allows all relevant physical parameters to be a generic function of the optical
depth: $T(\tau)$, $B(\tau)$,$\gamma(\tau)$, $\phi(\tau)$, $V_{\rm los}(\tau)$, etc. 
SIR retrieves the values of those parameters at a number of optical depth points called nodes. 
The final stratification is obtained by interpolating splines across those nodes. 
Given the high-spatial resolution of our observations we assume that the penumbral 
structure is horizontally resolved, thus we perform an inversion with only one magnetic component,
for which we allow 5 nodes in $T(\tau)$, 2 for $B(\tau)$, $\gamma(\tau)$, $\phi(\tau)$ 
and $V_{\rm los}(\tau)$, 1 for $V_{\rm mac}$ and $V_{\rm mic}$ 
(macro and microturbulent velocities). Including an extra free parameter to account for the 
scattered light, we have a total of 16 free parameters. The same scattered light profile is used in all 
inverted penumbral pixels. It is obtained by averaging the intensity profiles of those pixels with polarization 
signals below the noise level. The inversion retrieves typical values for the amount of scattered light of 8 to 17 \%.

As a result of applying the SIR inversion code, we obtain the magnetic field vector in the 
observer's reference frame at each spatial pixel as a function of the optical depth.
\footnote{An inversion of Hinode data from a different sunspot has been presented by Bellot 
Rubio et al. 2007 and Ichimoto et al. 2007. They considered only the 
Milne-Eddington case in which physical parameters 
are assumed to be height-independent.} To facilitate the interpretation of the results it 
is convenient to transform to the local reference frame. In this paper we will discuss 
only those results obtained along the inclined dashed lines in Fig.~1 (denoted as Y$^{*}$).
This region lies close to the line-of-symmetry of the sunspot (the line that connects the center 
of the sunspot with the center of the solar disk). In this region the conversion to the local 
reference frame, X$^{*}$Y$^{*}$Z$^{*}$, can be done as follows:

\begin{eqnarray}
B_x^{*}(\tau) & = & B (\tau) [ \sin \gamma(\tau) \cos \phi(\tau) \cos \theta - \cos \gamma(\tau) \sin \theta]\\
B_y^{*}(\tau) & = & B (\tau) \sin \gamma(\tau) \sin \phi(\tau)\\
B_z^{*}(\tau) & = & B (\tau) [\cos \gamma(\tau) \cos \theta + \sin \gamma(\tau) \cos \phi(\tau) \sin \theta]
\end{eqnarray}

\noindent where $\theta$ corresponds to the heliocentric angle of the sunspot and $B(\tau)$, $\gamma(\tau)$ 
and $\phi(\tau)$ are obtained from the inversion. Note that the knowledge of $\phi(\tau)$ is
affected by the inherent 180 degrees ambiguity. In order to distinguish between the two possible
solutions: $\phi$ and $\phi+\pi$, we consider that the magnetic field must point outwards from the sunspot center
and therefore $B_x^{*} > 0$. This allows one to uniquely determine the azimuthal angle. Finally, the inclination
and azimuth of the magnetic field vector in the local reference frame may be obtained through:

\begin{eqnarray}
\Psi & = & \tan^{-1} \left[\frac{B_y^{*}}{B_x^{*}}\right] \\
\zeta & = & \cos^{-1} \left[\frac{B_z^{*}}{B}\right]
\end{eqnarray}

\section{Results and discussion}

Figures 2 and 3 shows the vertical distribution of the line-of-sight velocity (upper left panel), 
magnetic field strength (upper right panel), magnetic field inclination $\zeta$ 
(lower left panel) and magnetic field azimuth $\Psi$ (lower right panel), along the left and right 
slices Y$^{*}$ respectively (dashed lines in Fig.~1). {\it Spinal} and {\it intraspinal} structures are clearly
visible. These examples display 6 full intraspines, denoted as {\it i1} through {\it i6} in the velocity maps.
Their vertical extension is about 1-1.5 logarithmic units of optical depth (roughly 150-250 Km). The horizontal 
extension shows larger variations, ranging from 300 (i2) to 1000 Km (i1). This is consistent  with the 
large pixel-to-pixel variations in the filling factor between the strong/vertical and weak/horizontal 
components obtained by Bellot Rubio et al. (2004), using a two-component inversion of spectropolarimetric 
observations at 1" resolution. From the inversion of other regions in the penumbra (not shown here) we also
see examples of some very large intraspines that appear to be formed by two smaller ones very  close to each other.

The magnetic field shows a distinct pattern in azimuthal angle. 
Above the intraspines $\Psi$ changes sign, being negative on the upper-left
 region above the intraspines, but positive on its right. This effect is highlighted
by the thick solid line in the maps of the line-of-sight velocity, field inclination and azimuth,
which show the variations of $\Psi$ at an optical depth level of $\tau=0.1$.
This effect unmistakably denotes a change in the sign of $B_y^{*}$ and therefore indicates that the spinal magnetic field wraps 
around the intraspine.  This is further confirmed by over-plotting the vector field 
(see arrows in Fig~2 and 3) in the Y$^{*}\tau$ plane \footnote{We do not convert to 
vertical scale Z$^{*}$ from the optical depth scale $\tau$ since this requires additional
assumptions (i.e.: hydrostatic equilibrium) that are not fully justified.}. 
Note that $B_x^{*}$ is missing in these figures, therefore the real magnetic field vectors 
will be pointing outside of that plane.

The wrapping effect is seen in four of the six presented examples. Above the 
two smaller intraspines (i2 and i5) $\Psi$ is negative on both sides, although it comes 
close to being positive on the right. This might indicate that horizontally narrow intraspines
perturb the surrounding field somewhat less than thick intraspines, where the external field is strongly 
forced to bend around them.

Lites et al. (1993) also found that $\Psi$ has opposite signs on both sides of the intraspines.
They carried out a Milne-Eddington inversion of ASP data at lower spatial resolution. 
In our study we used a more sophisticated inversion technique that accounts
for the optical depth dependence of the physical parameters. This kind
of inversion is particularly appropriate in the limb side
of the penumbra, where highly asymmetric Stokes profiles (oftentimes multi-lobed
Stokes V) are clear indicative of gradients along the line-of-sight in the magnetic 
field and velocity. As a result, we have also detected that the magnetic field in the spines
sometimes avoids the intraspines. This is the case of i2 and i6, 
where at optical-depth levels $\tau_5 \in [1,0.1]$, the magnetic  field has 
very small or no $B_y^{*}$ component: $\Psi \simeq 0$ and therefore $B \simeq B_x{*}$ 
(magnetic field inside the intraspines is aligned with the radial direction in the penumbra).
In the rest of the examples: i1, i3, i4 and i5, due to the strong non-vanishing $B_y^{*}$ component,
the magnetic field inside the intraspines does not point radially outwards from the sunspot.
Borrero (2007) has presented a magnetohydrostatic model for cylindrical flux tubes where the magnetic field 
is not aligned with the tube's axis. Although our results cannot be used to confirm or 
rule out that model, an analysis targeting this particular effect, could be more conclusive.

The Evershed flow is mainly concentrated in the intraspines, with velocities reaching up 
to 4.5 km s$^{-1}$. Velocities towards the observer (blueshifted velocities; $V_{\rm los} <0$) appear 
sometimes high above the intraspines. This feature could bear important similarities to the
upflows seen above umbral dots in 3D MHD simulations (Sch\"ussler \& V\"ogler 2006), however that
possibility is ruled out by the fact that inversions of our observations in the center-side
reveal red-shifted velocities in the top layers of the intraspines (which are now characterized by 
blue-shifts in their lower layers). Therefore this pattern seems more compatible 
with the Evershed \& inverse Evershed flow (Deming et al. 1988; see also Fig.~11 in
 Bellot Rubio et al. 2006). Another possibility is that this effect is only an artifact of 
the inversion process. In deep layers we observe strong red-shifted velocities, whose magnitude 
decays rapidly as we move towards higher layers. Since we only allow for 2 nodes in velocity 
(linear behavior with optical depth) the resulting velocities at $\tau_5 \sim 10^{-3}$ could 
be blue-shifted simply because no other option is allowed to the inversion code. We have 
repeated our inversions using 3 and 4 nodes and upflows, although weaker in magnitude, could 
still be seen above the intraspines. In the light of these results we cannot
draw a decisive conclusion, but it is certainly something worth studying in the future.

\begin{figure*}
\begin{center}
\begin{tabular}{cc}
\includegraphics[width=7cm]{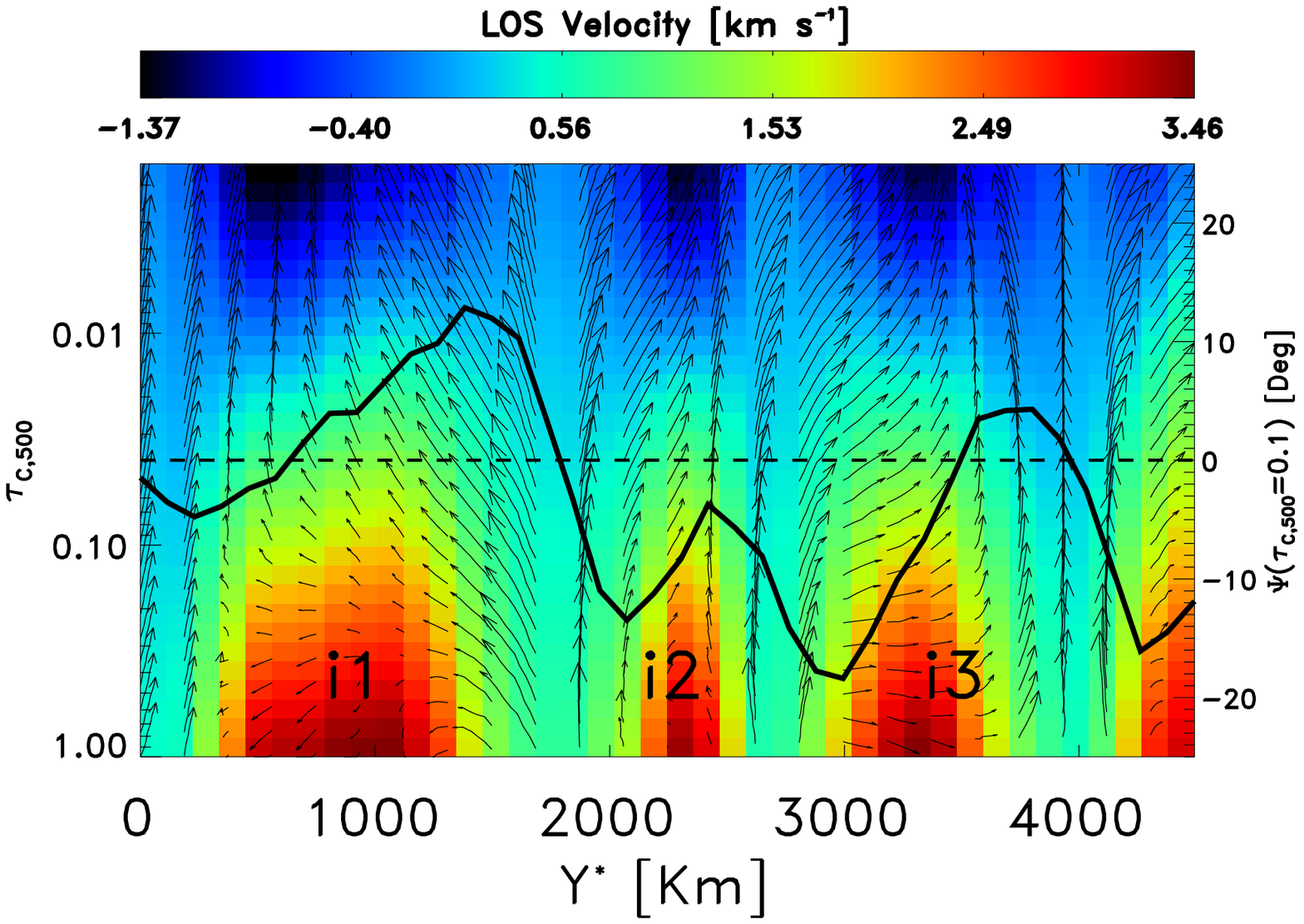} &
\includegraphics[width=7cm]{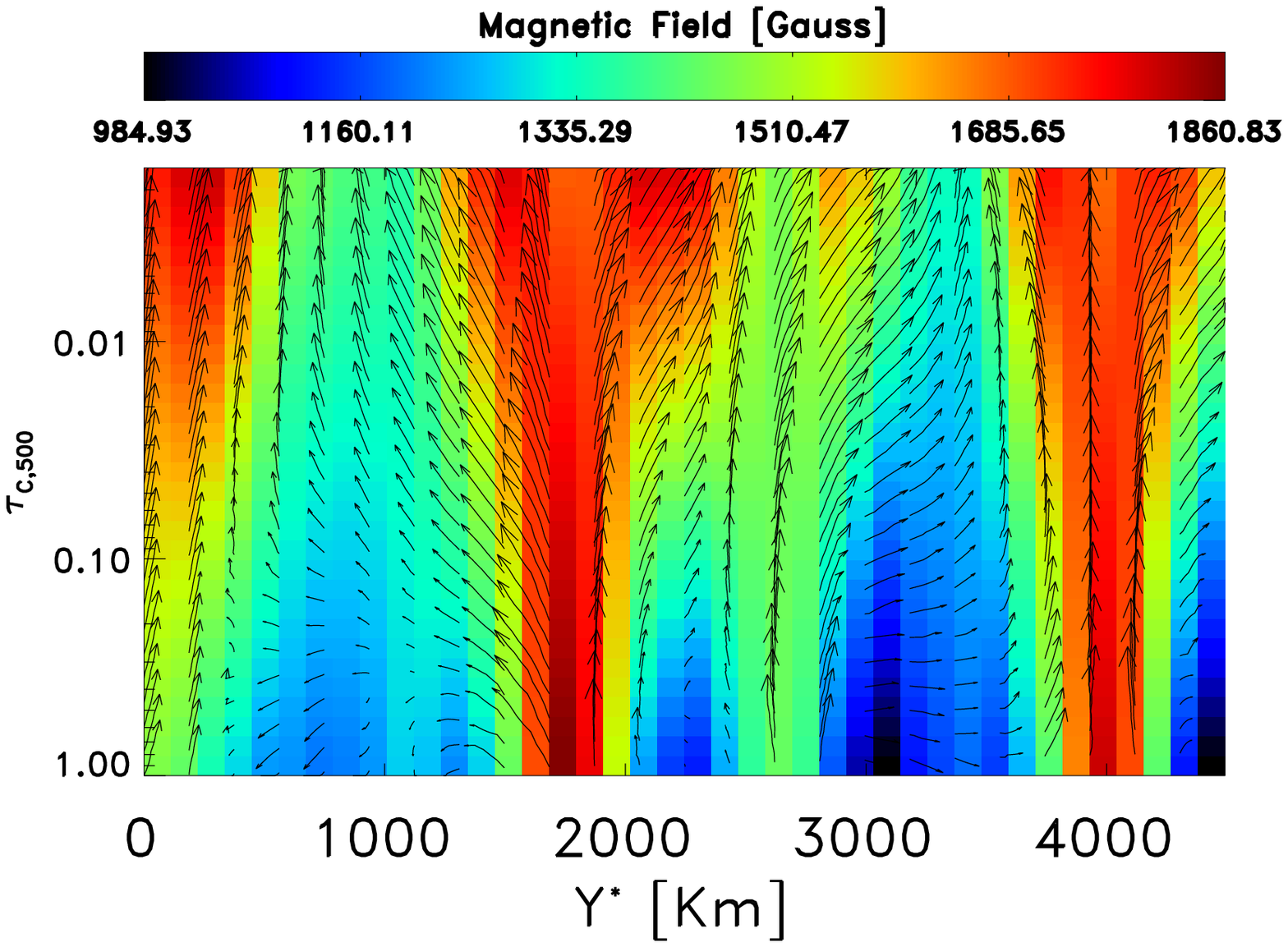} \\
\includegraphics[width=7cm]{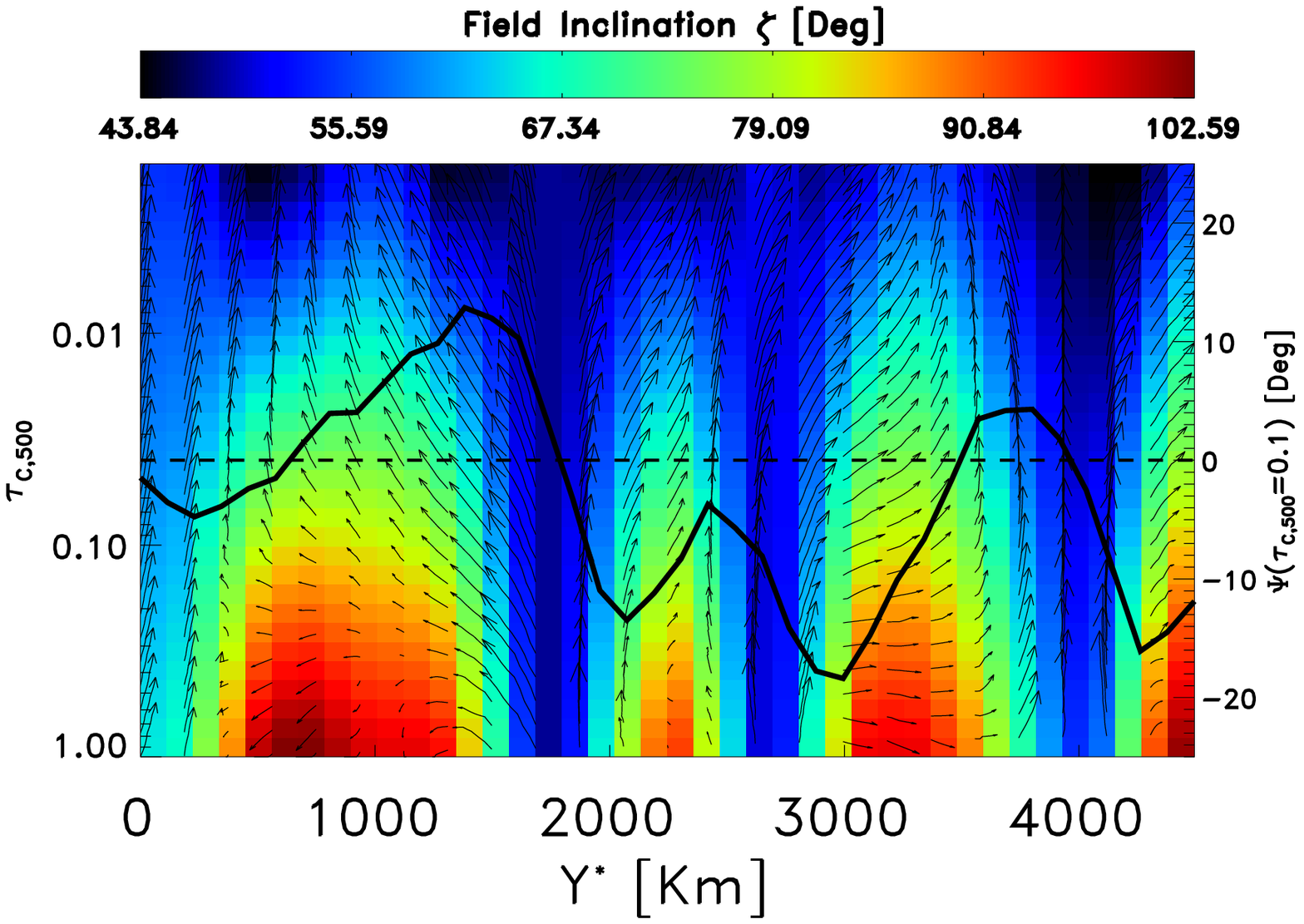} &
\includegraphics[width=7cm]{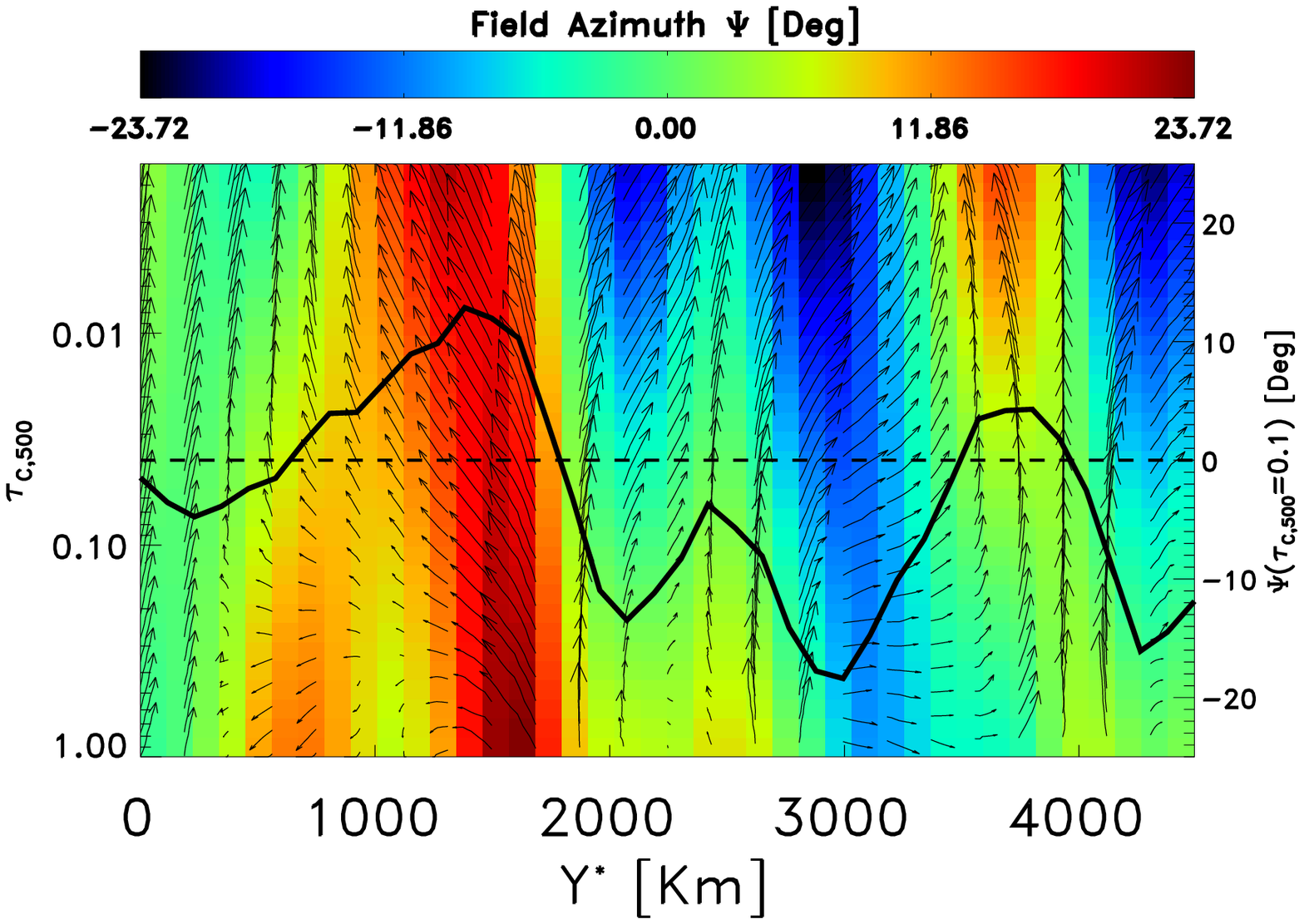}
\end{tabular}
\caption{Two dimensional cut across the leftmost Y$^{*}$ slice in Fig.~1 showing the depth-dependence of the
line of sight velocity (upper-left panel), magnetic field strength (upper-right panel), 
magnetic field inclination in the local reference frame $\zeta$ (lower-left panel) and
magnetic field azimuth in the local reference frame $\Psi$ (lower-right panel). The
magnetic field vector at different locations in the Y$^{*} \tau$ plane is indicated by the arrows.}
\end{center}
\end{figure*}

\begin{figure*}
\begin{center}
\begin{tabular}{cc}
\includegraphics[width=7cm]{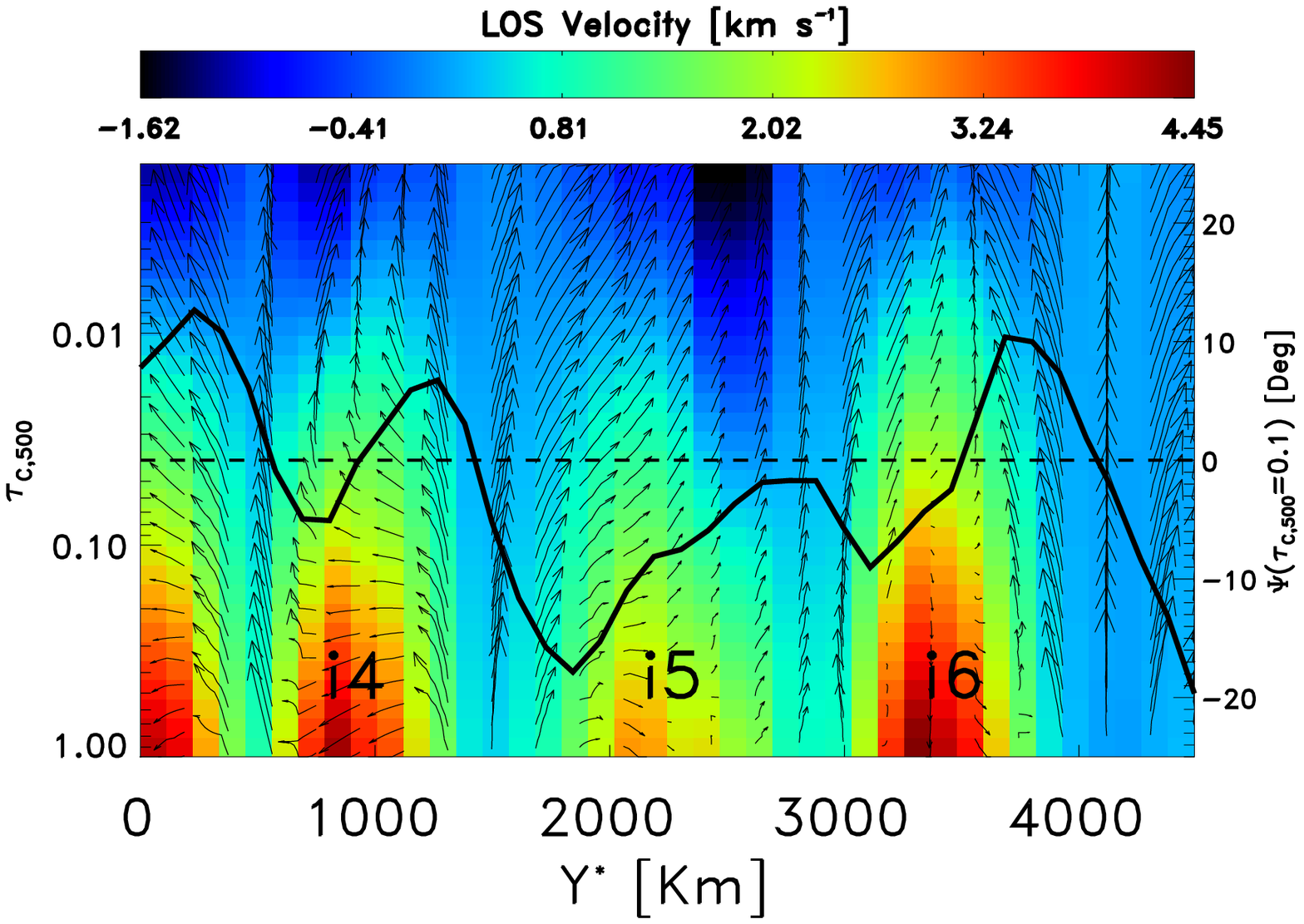} &
\includegraphics[width=7cm]{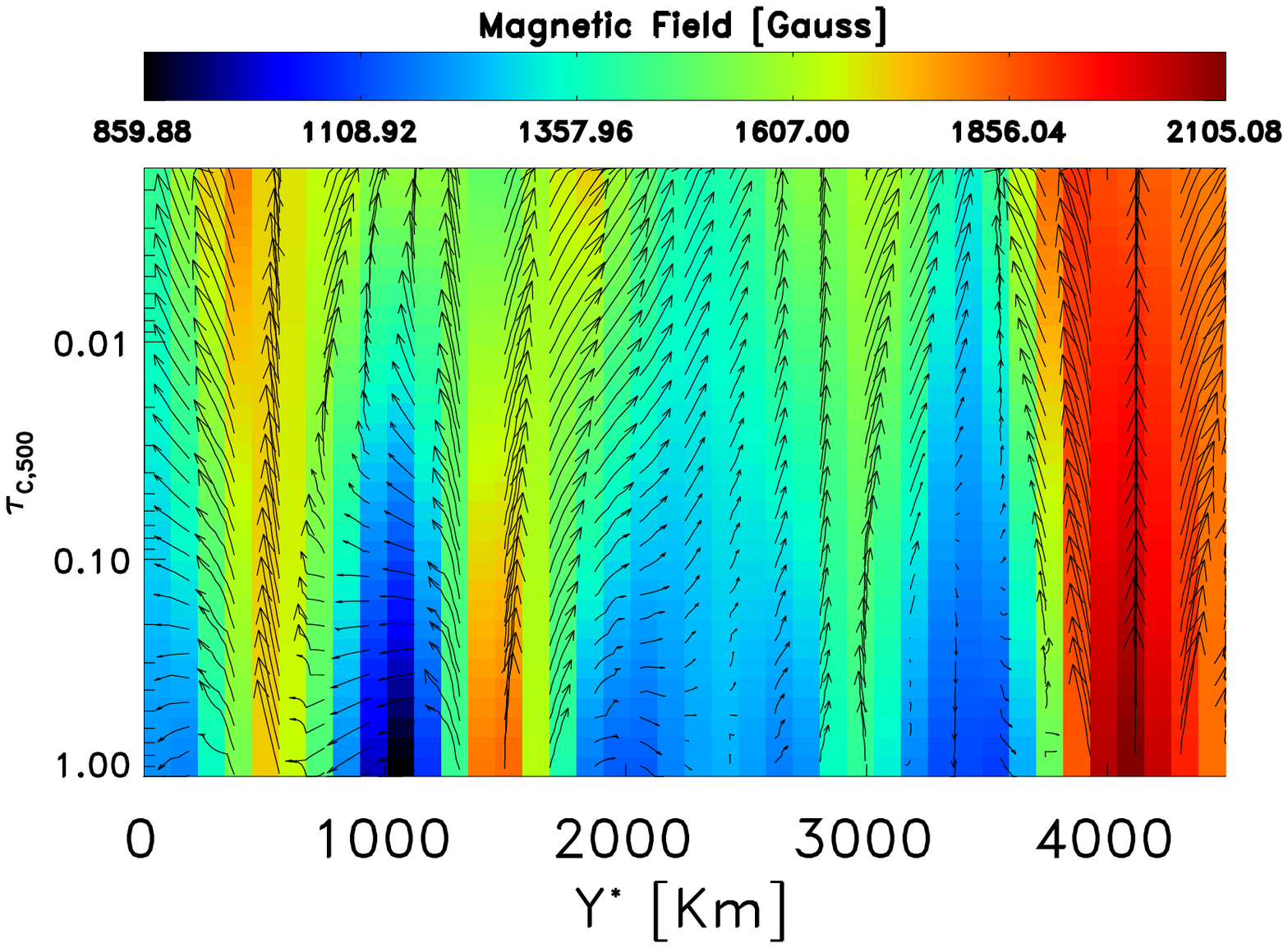} \\
\includegraphics[width=7cm]{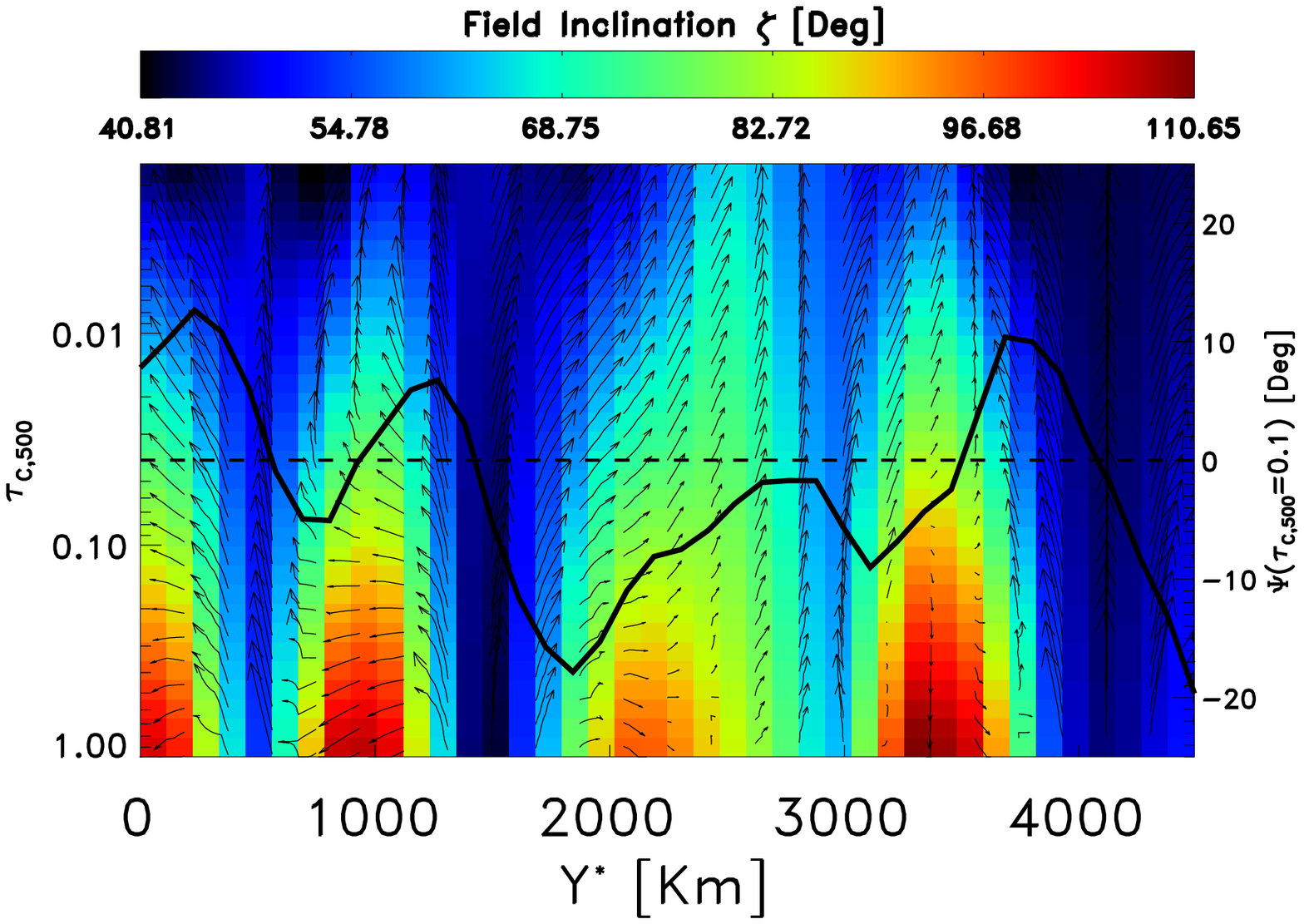} &
\includegraphics[width=7cm]{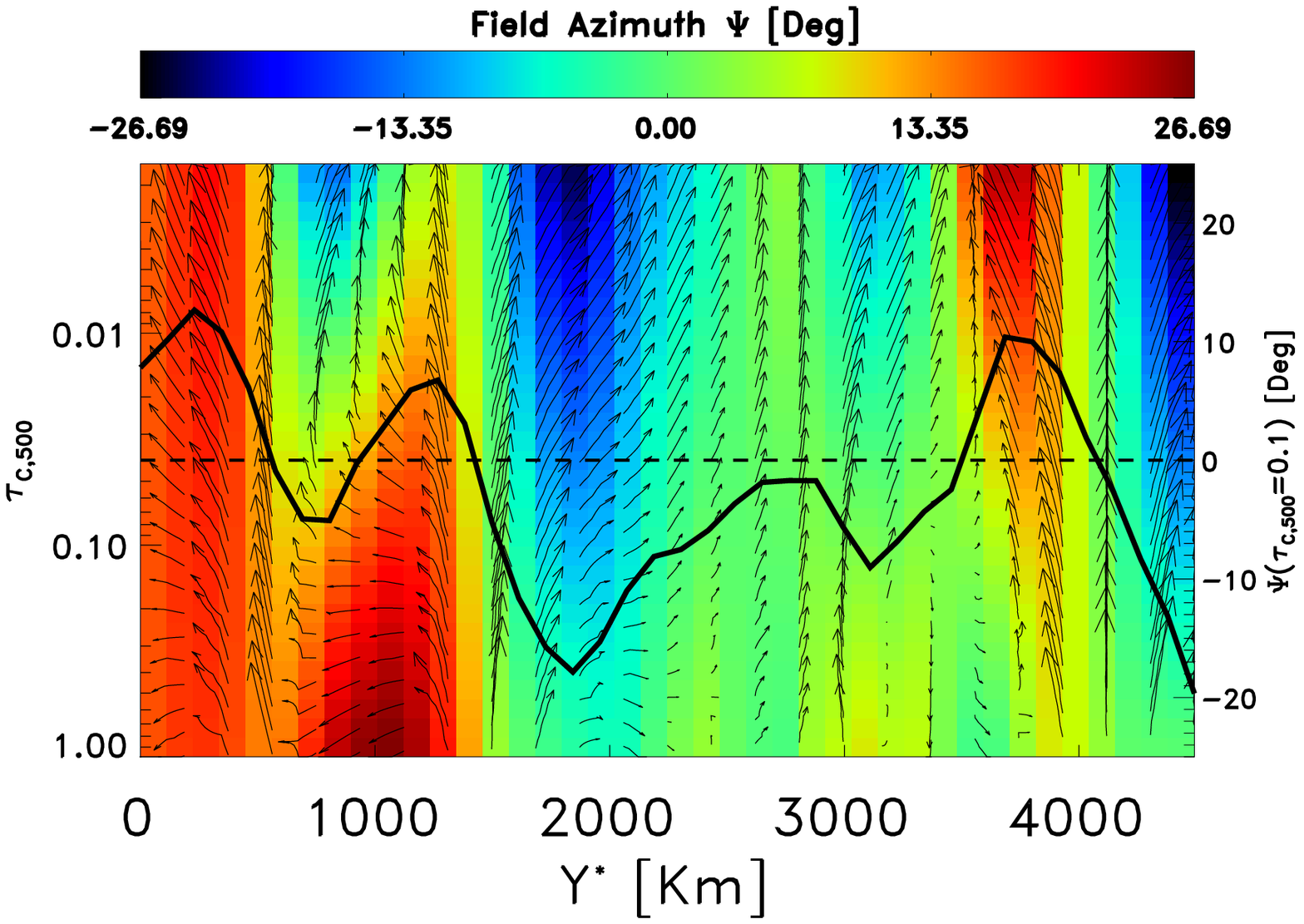}
\end{tabular}
\caption{Same as in Figure 2 but for the rightmost slice across Y$^{*}$ in Figure 1.}
\end{center}
\end{figure*}

\section{Conclusions}

Using high spatial resolution spectropolarimetric observations recorded by the
Hinode spacecraft, we have demonstrated for the first time that the magnetic field 
in the penumbral spines (strong and vertical magnetic field) folds and bends around 
the intraspines (weaker and more horizontal magnetic field) while avoiding them.
It is also frequent to observe how the transverse component of the external magnetic field 
leaks into the intraspines. The overall magnetic configuration is at odds with the MISMA model 
(S\'anchez Almeida 2005) for the penumbral magnetic field. However, it partially supports 
the geometry adopted by the uncombed model (Solanki \& Montavon 1993, 
cf. Borrero 2007) and the gappy model (Spruit \& Scharmer 2006).

Although at this point our results seem to match better with the uncombed penumbra, 
the agreement is not perfect. In addition, the highly simplified version 
of the gappy penumbra presented in Scharmer \& Spruit (2006) could be modified into a 
configuration similar to that of Sch\"ussler \& V\"ogler (2006), that presents more 
similarity to our findings. As our next step we will carry a more thorough analysis
of the differences and agreements between models and observations.

\begin{acknowledgements}
Thanks to the referee, Rolf Schlichenmaier, for his useful comments. In particular one that lead
to discover an error in our calculations.
\end{acknowledgements}

\end{document}